\documentclass[dvips,prd,twocolumn,
amssymb,nofootinbib,
letterpaper]{revtex4}
\usepackage[dvips]{graphicx}
\usepackage{subfigure}
\usepackage{setspace}
\usepackage{epsfig}
\usepackage{psfrag}
\begin{document}
\preprint{}

\title{Dirac neutrinos from a second Higgs doublet}

\author{Shainen M.\ Davidson}

\author{Heather E.\ Logan}
\email{logan@physics.carleton.ca}

\affiliation{Ottawa-Carleton Institute for Physics,
Carleton University, Ottawa K1S 5B6 Canada}

\date{October 15, 2009}

\begin{abstract}

We propose a minimal extension of the Standard Model in which
neutrinos are Dirac particles and their tiny masses are explained
without requiring tiny Yukawa couplings.  A second Higgs doublet with
a tiny vacuum expectation value provides neutrino masses while
simultaneously improving the naturalness of the model by allowing a
heavier Standard Model-like Higgs boson consistent with electroweak
precision data.  The model predicts a $\mu \to e \gamma$ rate
potentially detectable in the current round of experiments, as well as
distinctive signatures in the production and decay of the charged
Higgs $H^+$ of the second doublet which can be tested at future
colliders.  Neutrinoless double beta decay is absent.

\end{abstract}

\maketitle
\section{Introduction}

Since the discovery of neutrino oscillations, many models of neutrino
mass have been constructed.  The most straightforward is to
incorporate Dirac neutrino masses into the Standard Model (SM) by
introducing three (or two~\cite{Davoudiasl:2004be}) right-handed
neutrinos $\nu_R$ coupled to the SM Higgs doublet $\Phi_1$ analogously
to the SM quarks and charged leptons.  The Yukawa Lagrangian becomes,
\begin{eqnarray}
  \mathcal{L}_{Yuk} &=& - y^d_{ij} \bar d_{R_i} \Phi_1^{\dagger} Q_{L_j} 
    - y^u_{ij} \bar u_{R_i} \tilde \Phi_1^{\dagger} Q_{L_j} \nonumber \\
   & & - y^{\ell}_{ij} \bar e_{R_i} \Phi_1^{\dagger} L_{L_j} 
    - y^{\nu}_{ij} \bar \nu_{R_i} \tilde \Phi_1^{\dagger} L_{L_j}
    + {\rm h.c.},
    \label{eq:smyuk}
\end{eqnarray}
where $\tilde \Phi_1 \equiv i \sigma_2 \Phi_1^*$ is the conjugate of the
Higgs doublet, $Q_L \equiv (u_L, d_L)^T$ and $L_L \equiv (\nu_L,
e_L)^T$ are the left-handed quark and lepton doublets, respectively,
and a sum over the generation indices $i,j$ is implied.

This implementation has two problems.  First, realistic Dirac neutrino
masses below $\sim 1$~eV require nine independent dimensionless Yukawa
couplings $|y^{\nu}_{ij}| \lesssim 10^{-11}$.  Second, the
right-handed neutrinos $\nu_R$ are uncharged under the SM gauge group,
so that Majorana mass terms $M_{ij} \nu_{R_i} \nu_{R_j}$ are allowed
by the gauge symmetry.  The Majorana mass terms can be eliminated by
imposing a global symmetry such as lepton number.  Alternatively, a
large Majorana mass for the right-handed neutrinos leads to naturally
light left-handed Majorana neutrinos via the type-1
seesaw~\cite{Type1seesaw}, $m_{\nu} \sim (y^{\nu}v_1)^2/M$, where $v_1
\simeq 246$~GeV is the SM Higgs vacuum expectation value (vev).
This possibility motivates the experimental search for neutrinoless
double beta decay, which can happen only if the neutrino is a Majorana
particle.  Most other neutrino mass models also yield Majorana
neutrinos.

In this paper we introduce a minimal model for \emph{Dirac} neutrino
masses that does not require tiny neutrino Yukawa couplings.  Our
motivation is to provide a viable, renormalizable model with minimal
new field content which appears entirely below the TeV scale.  The
smallness of the neutrino masses relative to those of the quarks and
charged leptons is explained by sourcing them from a second Higgs
doublet $\Phi_2$ with a tiny vev $v_2 \sim$~eV.  The second Higgs
doublet yields two neutral scalars and a charged scalar pair at the
electroweak scale, providing signatures at the CERN Large Hadron
Collider (LHC) that can be used to discriminate the model from other
neutrino mass models and to extract new information about the neutrino
masses.  The charged scalar also contributes to the lepton flavor
violating decay $\mu \to e \gamma$ at a rate potentially within reach
of the currently-running MEG experiment, which would provide
additional sensitivity to the model parameters.  The model can be made
consistent with all existing experimental constraints, including
standard big bang nucleosynthesis (BBN) which generally constrains models
with new light degrees of freedom.  The second doublet has the
additional benefit of allowing a heavier SM-like Higgs boson to be
consistent with the precision electroweak data, thereby easing the
fine-tuning problem of a light Higgs boson~\cite{Barbieri:2006dq}.

In the next two sections we describe the model and show that
consistency with BBN can be achieved by keeping the neutrino Yukawa
couplings $y_i^{\nu}$ below about $1/30$.  In Sec.~\ref{sec:pheno} we
discuss the phenomenology.  After considering the decay modes of the
new Higgs bosons, we derive a constraint on the charged Higgs mass
from existing data from the CERN Large Electron-Positron (LEP)
collider.  We then address predictions for $\mu \to e \gamma$.  We
also show that the model is consistent with constraints from the muon
anomalous magnetic moment and tree-level muon and tau decay.  We
finish with a discussion of LHC search prospects for the charged Higgs
and possible effects on the phenomenology of the SM-like Higgs.
Finally we summarize our conclusions in Sec.~\ref{sec:conclusions}.

\section{The model}

The field content is that of the SM with the addition of a new scalar
doublet $\Phi_2$---with the same gauge quantum numbers as the SM Higgs
doublet $\Phi_1$---and three gauge-singlet right-handed neutrino
fields $\nu_{R_i}$ which will pair up with the three left-handed
neutrinos of the SM to form Dirac particles.  We impose a global U(1)
symmetry under which the new fields $\Phi_2$ and $\nu_{R_i}$ carry
charge $+1$ while all SM fields are uncharged.  This U(1) symmetry is
needed to forbid Majorana mass terms for the $\nu_{R_i}$ while
simultaneously enforcing a Yukawa coupling structure in which only
$\Phi_2$ couples to right-handed neutrinos.  The 4th term
in Eq.~\ref{eq:smyuk} is then replaced according to
\begin{equation}
  - y^{\nu}_{ij} \bar \nu_{R_i} \tilde \Phi_1^{\dagger} L_{L_j}
 \ \ \rightarrow \ \
  - y^{\nu}_{ij} \bar \nu_{R_i} \tilde \Phi_2^{\dagger} L_{L_j}.
\end{equation}
If the U(1) symmetry is unbroken, $\Phi_2$ has zero vev and the
neutrinos are strictly massless~\cite{Fayet:1974fj}.  An identical
Yukawa structure can be obtained using a $Z_2$ symmetry, as in the
models of Refs.~\cite{Ma:2000cc,Gabriel:2006ns}; however, this does
not forbid Majorana mass terms for $\nu_{R_i}$.  

In order to generate a vev for $\Phi_2$, we break the global U(1)
\emph{explicitly} using a dimension-2 term in the Higgs potential of
the form $m_{12}^2 \Phi_1^{\dagger} \Phi_2$.  This results in a
seesaw-like relation~\cite{Ma:2000cc}
\begin{equation}
   v_2 = m^2_{12} v_1/M_A^2,
\end{equation}
where $M_A$ is the mass of the neutral pseudoscalar (defined below).
For $M_A \sim 100$~GeV, $v_2 \sim$~eV is achieved for $m_{12}^2$ of
order (a few hundred keV)$^2$.  By breaking the global U(1) explicitly
we avoid an extremely light scalar as is present in the model of
Ref.~\cite{Gabriel:2006ns}; this allows us to satisfy BBN constraints
without resorting to nonstandard cosmology.

With these considerations the scalar potential is,
\begin{eqnarray}
  V &=& m_{11}^2 \Phi_1^{\dagger} \Phi_1 + m_{22}^2 \Phi_2^{\dagger} \Phi_2
  - [m_{12}^2 \Phi_1^{\dagger} \Phi_2 + {\rm h.c.}] 
  \nonumber \\
  && + \frac{\lambda_1}{2} ( \Phi_1^{\dagger} \Phi_1 )^2
  + \frac{\lambda_2}{2} ( \Phi_2^{\dagger} \Phi_2 )^2
  + \lambda_3 (\Phi_1^{\dagger} \Phi_1) (\Phi_2^{\dagger} \Phi_2) \nonumber \\
  && + \lambda_4 (\Phi_1^{\dagger} \Phi_2) (\Phi_2^{\dagger} \Phi_1).
  \label{eq:potential}
\end{eqnarray}
We can choose $m_{12}^2$ real and positive without loss of generality
by rephasing $\Phi_2$ and putting the excess phase into
$y^{\nu}_{ij}$, which is already a general complex matrix.  Stability
of the potential requires $\lambda_1,\lambda_2 > 0$, $\lambda_3 >
-\sqrt{\lambda_1 \lambda_2}$, and $\lambda_4 > -\sqrt{\lambda_1
\lambda_2} - \lambda_3$.  Note that even after the global U(1) is
broken, conventional lepton number survives as an accidental symmetry
of the model.  Neutrinoless double beta decay is thus
absent.\footnote{As in the SM, higher-dimensional operators can
violate lepton number; we assume such operators are Planck-suppressed.
The leading contribution is from $(L_L \Phi_1^{\dagger})^2/M_{Pl}$, which
yields a Majorana mass term for the left-handed neutrinos of order $3
\times 10^{-3}$~meV---compare effective electron neutrino Majorana
masses $|m_{ee}| > 14$~meV for the inverted neutrino mass hierarchy
and $|m_{ee}| > 2$~meV for the normal hierarchy (except for a possible
cancellation region) if the neutrinos are Majorana
particles~\cite{Aalseth:2004hb}.  Because the rate for neutrinoless
double beta decay is proportional to the square of the effective
electron neutrino Majorana mass, a Planck-suppressed Majorana mass term
would yield a neutrinoless double beta decay rate of order $10^6$
times smaller than the ultimate reach of the 100-ton-scale
experiments~\cite{Aalseth:2004hb} proposed to probe the normal
hierarchy in Majorana neutrino scenarios.}

The mass-squared parameters $m_{11}^2$ and $m_{22}^2$ suffer from
large radiative corrections with quadratic sensitivity to the
high-scale cutoff of the theory, just as the SM Higgs mass-squared
parameter does.  The resulting hierarchy problem could be solved as in
the SM by embedding our model into a supersymmetric or strong-dynamics
theory at the TeV scale.  Note however that because $m_{12}^2$ is the
\emph{only} source of breaking of the global U(1), its size is
technically natural---radiative corrections to $m_{12}^2$ are
proportional to $m_{12}^2$ itself and are only logarithmically
sensitive to the cutoff.  The smallness of $m_{12}^2$ required in our
model could thus be explained by higher-scale physics; e.g., through
spontaneous breaking of the U(1) in a hidden sector which is then
communicated to the Higgs sector by heavy messenger particles or at
high loop order.

We minimize the potential with the following considerations.  $\Phi_1$
must obtain the usual SM Higgs vev through spontaneous symmetry
breaking with $m_{11}^2 < 0$; neglecting $m_{12}^2$ and $v_2$ we
obtain $v_1^2 = -2m_{11}^2/\lambda_1$.  To avoid a
pseudo--Nambu-Goldstone boson with mass $\sim v_2$, the global U(1) is
\emph{not} also broken spontaneously; this is achieved for
$m_{22}^2 + (\lambda_3 + \lambda_4) v_1^2/2 > 0$.
Defining $\Phi_i = (\phi_i^+, (v_i + \phi_i^{0,r} + i
\phi_i^{0,i})/\sqrt{2})^T$ and neglecting mixing terms suppressed by
$v_2/v_1$, the mass eigenstates are two neutral scalars $h^0 \simeq
\phi_1^{0,r}$ (SM-like) and $H^0 \simeq -\phi_2^{0,r}$, a charged
scalar $H^+ \simeq -\phi_2^+$, and a neutral pseudoscalar $A^0 \simeq
-\phi_2^{0,i}$.  Mixing between $\Phi_1$ and $\Phi_2$ can be ignored
when $y^{\nu}_i \gg v_2/v_1$, as we assume here.  The physical masses
are,
\begin{eqnarray} 
  && M_h^2 = \lambda_1 v_1^2, \qquad \qquad
  M_{H^+}^2 = m_{22}^2 + \lambda_3 v_1^2/2, \nonumber \\ 
  && M_{A,H}^2 = M_{H^+}^2 + \lambda_4 v_1^2/2,
\label{eq:masses}
\end{eqnarray} 
where we again neglect terms suppressed by $v_2/v_1$.  In particular,
$H^0$ and $A^0$ are degenerate and can be lighter or heavier than
$H^+$ depending on the sign of $\lambda_4$.  Note that a mass
splitting $(M_{H^+} - M_A)$ of either sign in this model yields a
positive contribution to the $\rho$ parameter~\cite{Haber:1992cn},
which serves to increase the best-fit value for the SM-like Higgs mass
$M_h$ in the electroweak fit.  This eases the
tension~\cite{Chanowitz:2001bv} between the standard electroweak fit,
which prefers a light Higgs in the SM, and the lower bound on the SM
Higgs mass from LEP.  It also eases the ``little hierarchy problem,''
i.e., the fine-tuning of the Higgs mass-squared parameter against
radiative corrections required for a cutoff above
1~TeV~\cite{Barbieri:2006dq}.

In the limit $v_2 \ll v_1$ the Yukawa couplings of the new scalars are
given by
\begin{eqnarray}
\mathcal{L}_{Yuk} &=& (m_{\nu_i}/v_2) H^0 \bar \nu_i \nu_i
  - (i m_{\nu_i}/v_2) A^0 \bar \nu_i \gamma_5 \nu_i \nonumber \\
  & & - (\sqrt{2} m_{\nu_i}/v_2) [U_{\ell i}^* H^+ \bar \nu_i P_L e_{\ell}
    + {\rm h.c.}],
\end{eqnarray}
where $U_{\ell i}$ is the Pontecorvo-Maki-Nakagawa-Sakata (PMNS)
matrix, for which we use the convention of Ref.~\cite{Fogli:2005cq}.
In particular, the scalar Yukawa couplings are entirely fixed by
neutrino-sector parameters and the vev $v_2$ of the second doublet.

The scalars also couple among themselves; neglecting couplings
suppressed by $v_2$, the Feynman rules for all triple-scalar couplings
in the model are $-i \lambda_3 v_1$ for the $h^0 H^+ H^-$ vertex, $-i
(\lambda_3 + \lambda_4) v_1$ for the $h^0 H^0 H^0$ and $h^0 A^0 A^0$
vertices, and $-3i \lambda_1 v_1$ for the $h^0 h^0 h^0$ vertex.

\section{Big bang nucleosynthesis}
\label{sec:bbn}

In our model the right-handed neutrino degrees of freedom will be
populated in the early universe due, e.g., to $\ell^+ \ell^-
\leftrightarrow \nu_R \bar \nu_R$ via t-channel $H^+$ exchange.  The
model is thus constrained by the limit on new relativistic degrees of
freedom during BBN, $\delta N_{\nu,max} = 1.44$~\cite{Cyburt:2004yc}
at 95\% confidence level (CL).  To evade this bound, the right-handed
neutrinos must be colder than the left-handed neutrinos,
$T_{\nu_R}/T_{\nu_L} \leq (\delta
N_{\nu,max}/3)^{1/4}$~\cite{Steigman:1979xp}.  This can be achieved if
the right-handed neutrinos drop out of thermal equilibrium early
enough; in terms of the effective number of relativistic degrees of
freedom $g_*$ at the times of decoupling of the left- and right-handed
neutrinos we have $T_{\nu_R}/T_{\nu_L} = (g_{*L}/g_{*R})^{1/3}$.
Inserting $g_{*L} = N_B + 7N_F/8 = 43/4$ (where $N_B = 2$ photon and
$N_F = 10$ $e^{\pm}$ and $\nu_L$ spin degrees of freedom), we find
$g_{*R} \geq 43/4 + 7.9$.  We therefore require that the $\nu_R$
decouple from the thermal bath above the quark-hadron transition at
200--400~MeV, yielding 51 extra effective degrees of freedom from
muons, $u$, $d$, and $s$ quarks, and gluons.  

This puts an upper bound
on the $H^+$-mediated $\ell^+ \ell^- \leftrightarrow \nu_R \bar \nu_R$
cross section via~\cite{Steigman:1979xp} $T_{d,\nu_R}/T_{d,\nu_L}
\approx (\sigma_R/\sigma_L)^{-1/3} = [ 4 v_2^4 M_{H^+}^4 / v_1^4
m_{\nu_i}^4 |U_{\ell i}|^4 ]^{1/3}$.  Using $T_{d,\nu_L} \simeq
3$~MeV~\cite{Olive:1999ij} and imposing $T_{d,\nu_R} \gtrsim 300$~MeV
we obtain an upper bound on the neutrino Yukawa couplings, 
\begin{equation}
  y^{\nu}_i \equiv \sqrt{2} m_{\nu_i}/v_2 \lesssim
  \frac{1}{30}\left[\frac{M_{H^+}}{100 \ {\rm GeV}}\right]
  \left[\frac{1/\sqrt{2}}{|U_{\ell i}|} \right], 
\end{equation}
which for $M_{H^+} \sim 100$~GeV is comparable to the SM bottom quark
Yukawa coupling.  If the lightest neutrino is nearly massless, we may
take, for the normal neutrino mass hierarchy, $m_{\nu_3} \sim
\sqrt{\Delta m^2_{32}} \sim 0.05$~eV and $U_{\mu 3} \simeq
1/\sqrt{2}$, which yields $v_2 \gtrsim 2$~eV.  In the inverted
hierarchy the limit due to $y^{\nu}_{1,2}$ is comparable.  The limit
on $v_2$ scales linearly with the heaviest neutrino mass.

\section{Phenomenology}
\label{sec:pheno}

We now consider the phenomenology of the scalar sector.  
The decays of the new scalars are controlled by
the underlying U(1) symmetry---for $M_{A,H} > M_{H^+}$ ($M_{A,H} <
M_{H^+}$), the decay modes are $H^+ \to \ell^+ \nu$ and $A^0, H^0 \to
\nu \bar\nu, W^{\pm}H^{\mp}$ ($H^+ \to \ell^+ \nu, W^+ A^0, W^+ H^0$
and $A^0,H^0 \to \nu \bar \nu$).  All other decays are suppressed by
the tiny U(1) breaking $v_2$.
In particular, the tree-level couplings of $H^0$, $A^0$ to $W$ and $Z$
bosons, quarks, and charged leptons are suppressed by $v_2/v_1$.
Decays to gluons (via a quark loop) are thus also suppressed.  The
decays $A^0$, $H^0 \to \gamma \gamma$ through an $H^+$ loop are
suppressed by the tiny $A^0H^+H^-$, $H^0H^+H^-$ couplings $\sim
\lambda_2 v_2$.  The decay $H^0 \to h^0 h^0$ is also suppressed by
$v_2$.  Loop-induced decays of $H^0 \to ZZ, WW$ or $H^+ \to W^+Z,
W^+\gamma$ through a lepton triangle are suppressed by a neutrino mass
insertion.

In what follows we assume $M_{A,H} \geq M_{H^+}$ (i.e., $\lambda_4
\geq 0$) and focus on the decays of the charged Higgs.  First, note
that $H^-$ decays via the neutrino Yukawa couplings into a
\emph{left-handed} charged lepton, in contrast to the usual Type-I or
II two Higgs doublet model (2HDM)~\cite{HHG} in which $H^-$ decays
into a right-handed charged lepton.  In particular, $H^- \to \tau^-
\bar \nu$ produces a left-handed $\tau^-$, so that usual charged Higgs
searches that take advantage of the $\tau$ helicity to suppress $W$
backgrounds are not applicable in this model.

The charged Higgs decays into all nine combinations of $\ell_i \nu_j$.
Summing over final-state neutrinos we obtain the decay width into a
particular charged lepton species $\ell$, 
\begin{equation}
  \Gamma \left(H^+ \to \ell^+ \nu \right) 
  = \frac{M_{H^+} \langle m^2_{\nu} \rangle_{\ell}}{8\pi v_2^2},
\end{equation}
where we define the expectation value of the neutrino mass-squared in
a flavor eigenstate, $\langle m^2_{\nu} \rangle_{\ell} = \sum_i
m^2_{\nu_{i}} |U_{\ell i}|^2$~\cite{Fukuyama:2008sz} (here $\langle
m_{\nu}^2 \rangle_e$ is the same observable that is measured in
tritium beta-decay endpoint experiments like
KATRIN~\cite{Osipowicz:2001sq}).  Imposing the BBN constraint
$y_i^{\nu} \lesssim 1/30$ yields an upper bound on the $H^+$ total
width, $\Gamma_{H^+}^{tot} \lesssim 1.3 \times 10^{-4} M_{H^+}$, or
for $M_{H^+} = 100$~GeV, $\Gamma_{H^+}^{tot} \lesssim 13$~MeV.  The
charged Higgs is thus narrow but not long-lived.
Similarly, the new neutral Higgs widths into neutrinos are given by
$\Gamma(H^0 \to \nu_i \bar \nu_i) = \Gamma(A^0 \to \nu_i \bar \nu_i) =
M_A m_{\nu_i}^2/8 \pi v_2^2$, yielding an upper bound on the width to
neutrinos of $\Gamma(H^0, A^0 \to \nu \bar\nu) \lesssim 6.6 \times
10^{-5} M_A$.

\begin{figure}
\resizebox{0.5\textwidth}{!}{
\rotatebox{270}{\includegraphics{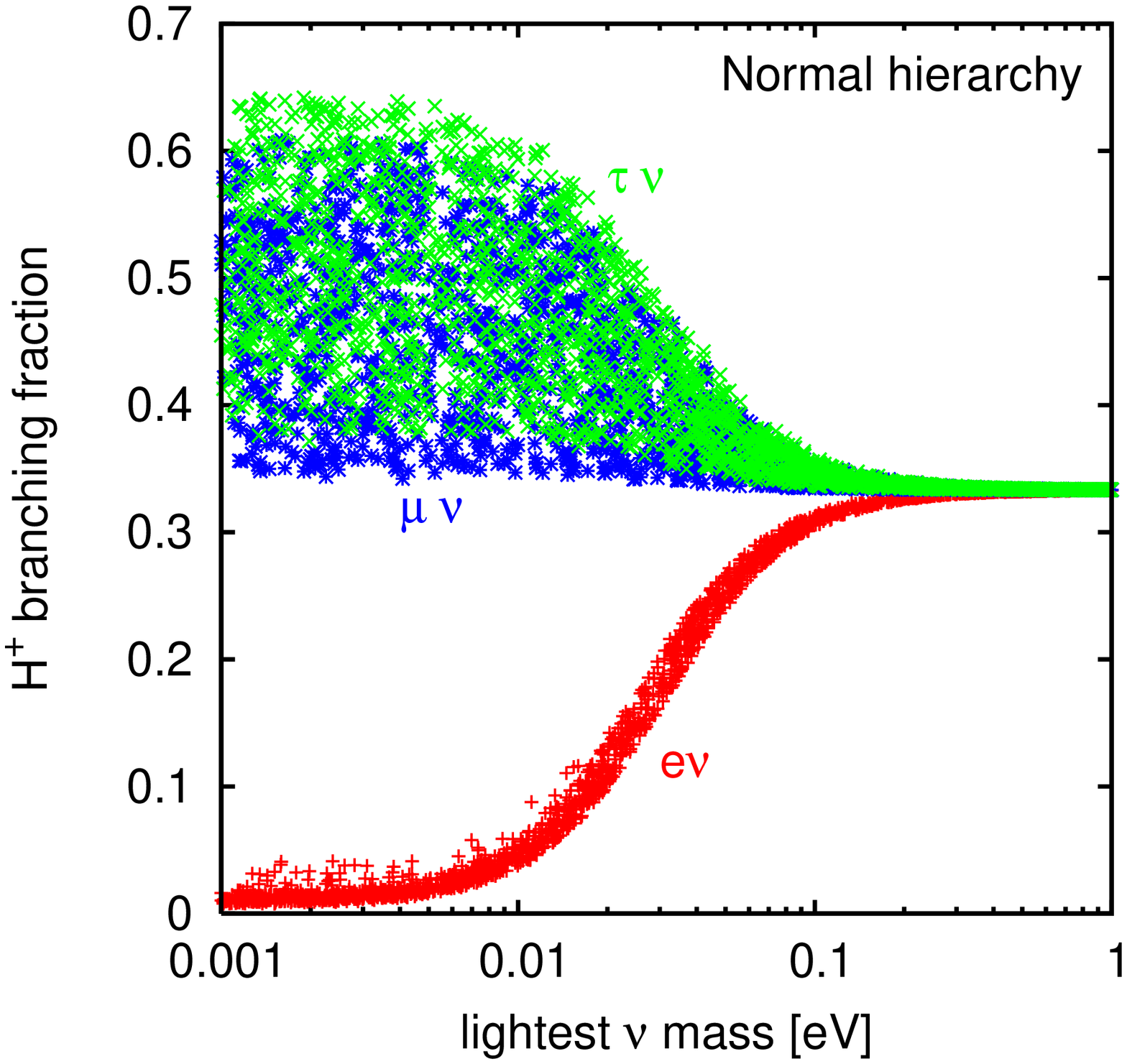}}
\rotatebox{270}{\includegraphics{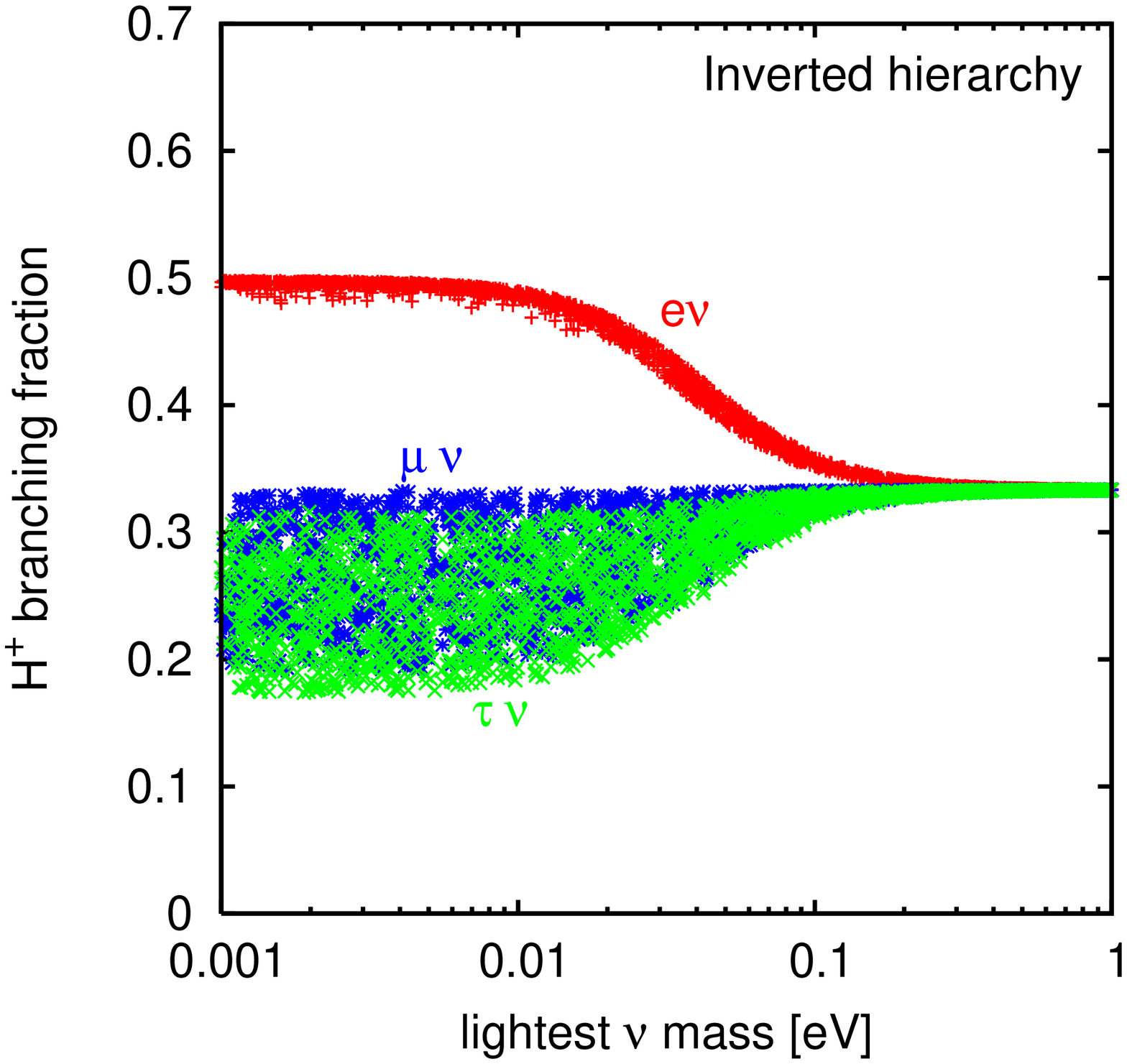}}}
\caption{Charged Higgs decay branching fractions to $e\nu$, $\mu\nu$,
and $\tau\nu$ as a function of the lightest neutrino mass.}
\label{fig:brs}
\end{figure}

The charged Higgs branching ratios are given by ${\rm BR}(H^+ \to
\ell^+ \nu) = \langle m^2_{\nu} \rangle_{\ell} / \sum_{\ell} \langle
m^2_{\nu} \rangle_{\ell}$ and are shown in Fig.~\ref{fig:brs} where we
scan over the 2$\sigma$ neutrino parameter ranges from
Ref.~\cite{Fogli:2005cq}.  These branching ratios are the same as
those of the singly-charged triplet Higgs state $\Phi^+$ in the type-2
seesaw model~\cite{Type2seesaw} as given in Ref.~\cite{Perez:2008ha}.

\subsection{Constraints from LEP}

Searches for leptons plus missing energy at LEP can be used to set
limits on the charged Higgs in this model.  Charged Higgs pair
production is dominantly through s-channel $Z$ and $\gamma$ exchange;
the t-channel neutrino exchange amplitude is suppressed by two powers
of $y^{\nu}_i \lesssim 1/30$ and we thus neglect it.  The usual LEP
charged Higgs search~\cite{LEPchargedHiggs} relies on $H^+$ decays to
$\tau\nu$ or $q \bar q^{\prime}$, as are expected in the Type-I
or II 2HDMs~\cite{HHG}.  In our model, charged Higgs decays to quarks
are absent and BR$(H^+ \to \tau^+\nu)$ reaches at most 0.65 for the
normal hierarchy and 0.33 for the inverted hierarchy
(Fig.~\ref{fig:brs}); because of the sizable branching fractions of
$H^{\pm}$ into $e \nu$ or $\mu \nu$, we find that these channels will
provide a stronger exclusion than the usual $\tau\nu$ channel.  

\begin{figure}
  \includegraphics*[width=0.5\textwidth]{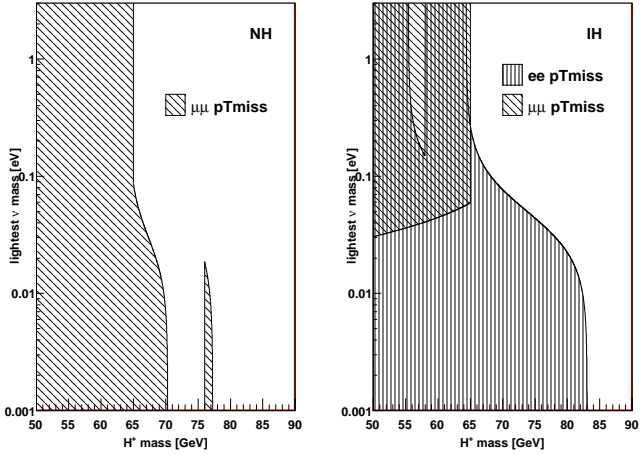}
     \caption{Charged Higgs mass lower bound as a function of the
lightest neutrino mass for the normal (NH) and inverted (IH) neutrino
mass hierarchies, based on LEP selectron and smuon cross section
limits from Ref.~\cite{lepsusy}.}
     \label{fig:mhlimit}
\end{figure}

LEP studied $e^+e^- p_T^{miss}$ and $\mu^+\mu^- p_T^{miss}$ in the
context of searches for the supersymmetric scalar partners of $e$ and
$\mu$ (selectrons and smuons) with decays to the corresponding lepton
plus a neutralino which escapes the
detector~\cite{lepsusyexp,lepsusy}.  For a massless neutralino the
kinematics reproduce those of $H^+H^-$ pair production.  Using the
smallest allowed branching fraction into the relevant decay mode from
Fig.~\ref{fig:brs}, we translate the LEP-combined 95\% CL cross
section limits~\cite{lepsusy} into a lower bound on $M_{H^+}$ as shown
in Fig.~\ref{fig:mhlimit}.  We find $M_{H^+} \gtrsim 65$--83 GeV,
depending on the hierarchy and the lightest neutrino mass.  These
bounds could be improved by a combined analysis of the $e^+ e^-
p_T^{miss}$, $\mu^+ \mu^- p_T^{miss}$, and $\tau^+ \tau^- p_T^{miss}$
channels and the inclusion of mixed-flavor channels.

For $M_A < M_{H^+}$, decays of $H^+ \to W^+ A^0, W^+ H^0$ could
compete with the leptonic decay and potentially invalidate our LEP
limits quoted above.  This bosonic decay proceeds via the SU(2) gauge
coupling; partial widths for offshell decays $H^+ \to W^+ A^0/H^0$
were computed in Ref.~\cite{Djouadi:1995gv} and are implemented in the
public FORTRAN code {\tt HDECAY}~\cite{HDECAY}.  We note however that
the $Z$ boson invisible width~\cite{Zpolereview} constrains $M_A \geq
43.7$~GeV at 95\% CL.  Taking $M_{H^+} = 83$~GeV, the largest possible
partial width for offshell $H^+ \to W^+ A^0/H^0$ relevant to the LEP
limits presented here is then 0.12~MeV.  Taking $y^{\nu}_i$ at the BBN
bound $\sim 1/30$, offshell decays constitute a branching ratio of
less than 1\%, so that our LEP limits remain valid.  The branching
ratio for offshell decays can reach 10\% for $y^{\nu}_i \sim 1/100$.

\subsection{$\mu \to e \gamma$ and $\tau \to \ell \gamma$}

\begin{figure}
\resizebox{0.5\textwidth}{!}{
\rotatebox{270}{\includegraphics{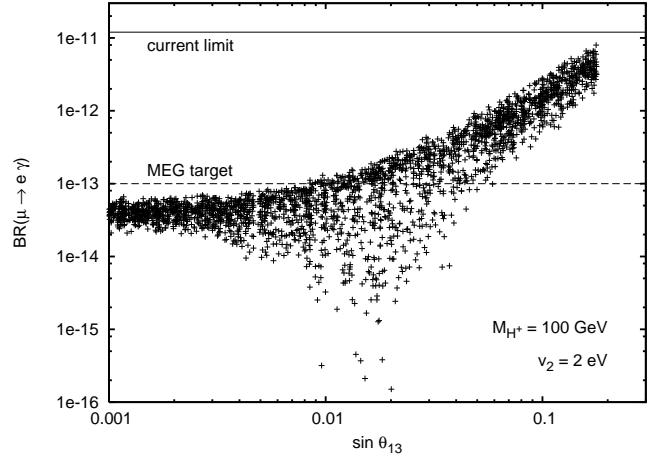}}}
\caption{Branching ratio of $\mu \to e \gamma$ versus
$\sin\theta_{13}$.  The normal and inverted hierarchies 
populate the same region.}
\label{fig:meg}
\end{figure}

At one loop, $H^+$ mediates lepton flavor violating decays $L \to
\ell \gamma$ with a branching
fraction~\cite{Nguyen:1987jg,Fukuyama:2008sz}
\begin{equation}
  {\rm BR}(L \to \ell \gamma) 
  = {\rm BR}(L \to e \nu \bar\nu) 
  \frac{\alpha_{em}}{96 \pi} 
  \frac{|\sum_i m_{\nu_i}^2 U_{L i}^* U_{\ell i} |^2}
       {2 G_F^2 M_{H^+}^4 v_2^4},
\end{equation}
where $(L,\ell) = (\mu, e)$ or $(\tau, e$ or $\mu)$.\footnote{The type-2
seesaw yields an analogous formula~\cite{Akeroyd:2009nu} with the same
dependence on the neutrino parameters but dominated by loops involving
the doubly-charged triplet state $\Phi^{++}$.}  Unitarity of the PMNS
matrix yields $\sum_i m_{\nu_i}^2 U_{L i}^* U_{\ell i} = - \Delta
m^2_{21} U_{L 1}^* U_{\ell 1} + \Delta m^2_{32} U_{L 3}^* U_{\ell 3}$.
Scanning over the allowed neutrino parameter
ranges~\cite{Fogli:2005cq} with $M_{H^+} = 100$~GeV and $v_2 = 2$~eV
yields BR($\mu \to e \gamma) \leq 8 \times 10^{-12}$, with significant
dependence on $\sin\theta_{13} \equiv |U_{e3}|$ as shown in
Fig.~\ref{fig:meg}.\footnote{Note that the rate for $\mu \to e \gamma$
($\tau \to e \gamma$) vanishes when $U_{e3} = [\Delta m^2_{21}/\Delta
m^2_{32}][U^*_{L1} U_{e1}/U^*_{L3}]$, with $L = \mu$
($\tau$)~\cite{Rodejohann:2008xp}.  The rate for $\tau \to \mu \gamma$
cannot vanish given the known neutrino parameter values.}  This
branching fraction is below the current experimental 90\% CL upper
limit of BR$(\mu \to e \gamma) \leq 1.2 \times
10^{-11}$~\cite{Brooks:1999pu}, but for $\sin\theta_{13} \gtrsim 0.01$
is within reach of the MEG experiment which expects a sensitivity of
about $10^{-13}$ after running to the end of 2011~\cite{MEGstatus}.
In particular, in the MEG sensitivity range the $\Delta m^2_{32}
U_{\mu 3}^* U_{e 3}$ term dominates, so that BR($\mu\to e\gamma)
\propto \sin^2 \theta_{13}/v_2^4 M_{H^+}^4$.  This provides a
measurement of $v_2$ if $\sin\theta_{13}$ ($M_{H^+}$) can be obtained
from neutrino oscillation (collider) experiments.

Similarly, for $M_{H^+} = 100$~GeV and $v_2 = 2$~eV we find BR($\tau
\to \mu \gamma) = (0.7$--$2.1) \times 10^{-11}$ and BR($\tau \to e
\gamma) \leq 1.5 \times 10^{-12}$.  These are well below the current
90\% CL bounds of $6.8 \times 10^{-8}$~\cite{Aubert:2005ye} and $1.1
\times 10^{-7}$~\cite{Aubert:2005wa} respectively, as well as the
expected reach at the SuperB next-generation flavor factory of $2
\times 10^{-9}$ in either channel~\cite{Bona:2007qt}.

\subsection{Muon anomalous magnetic moment}

The charged Higgs also contributes to the muon anomalous magnetic
moment $a_{\mu} \equiv (g - 2)/2$.  Adapting the calculation of
Ref.~\cite{Dedes:2001nx} we find the one-loop $H^+$ contribution,
\begin{equation}
   \delta a_{\mu}^{H^+} = - \frac{m_{\mu}^2 \langle m_{\nu}^2 \rangle_{\mu}}
   {48 \pi^2 M_{H^+}^2 v_2^2}.
\end{equation}  
Taking $M_{H^+} \sim 100$~GeV and $y_i^{\nu} \lesssim 1/30$ results in
$\delta a_{\mu}^{H^+}$ two to three orders of magnitude smaller than
the current experimental uncertainty on $a_{\mu}$.  The two-loop $H^+$
contribution~\cite{Grifols:1979yk} is also well below the current
sensitivity.

\subsection{Tree-level muon and tau decay}

In our model the decays $L \to \ell \nu \bar \nu$ receive
contributions from tree-level charged Higgs exchange.  Neglecting
$m_{\nu}$ in the kinematics, the differential cross section for the
$H^+$-mediated process is identical to that for the $W$-mediated
process but with the $\nu$ and $\bar\nu$ momenta interchanged, and the
interference term is zero.  The decay widths become $\Gamma(L \to \ell
\nu \bar \nu) = \Gamma^{SM}(L \to \ell \nu \bar \nu) [ 1 + \langle
m_{\nu}^2 \rangle_L \langle m_{\nu}^2 \rangle_{\ell} / 8 G_F^2
M_{H^+}^4 v_2^4 ]$~\cite{Fukuyama:2008sz}, where the neutrino mass
dependence in the $H^+$ contribution violates lepton flavor
universality.  In the standard parameterization (see, e.g.,
Ref.~\cite{Roney:2007zz}) we obtain $g_{\mu}/g_e \simeq 1 + \langle
m_{\nu}^2 \rangle_{\tau} ( \langle m_{\nu}^2 \rangle_{\mu} - \langle
m_{\nu}^2 \rangle_e) / 16 G_F^2 M_{H^+}^4 v_2^4$ and $g_{\mu}/g_{\tau}
\simeq 1 + \langle m_{\nu}^2 \rangle_e ( \langle m_{\nu}^2
\rangle_{\mu} - \langle m_{\nu}^2 \rangle_{\tau}) / 16 G_F^2 M_{H^+}^4
v_2^4$.  Taking $M_{H^+} \simeq 100$~GeV and imposing the BBN
constraint $y_i^{\nu} \lesssim 1/30$ results in deviations at the
$10^{-6}$ level, well within the current experimental constraints
$g_\mu/g_\tau = 0.9982 \pm 0.0021$ and $g_\mu/g_e = 0.9999 \pm
0.0020$~\cite{Roney:2007zz} as well as the expected SuperB reach of
$\pm 0.0005$ in either quantity~\cite{Roney:2007zz}.

\subsection{$H^+H^-$ production at LHC}

We now consider charged Higgs search prospects at the LHC.  The
charged Higgs can be pair produced via $pp \to \gamma^*,Z^* \to
H^+H^-$; we compute the cross section including next-to-leading-order
QCD corrections using {\tt
PROSPINO}~\cite{Beenakker:1999xh,Alves:2005kr}.  For $M_{H^+} = 100$
(500)~GeV we find a cross section of 300 (0.60)~fb, with a theoretical
uncertainty of $\sim$25\%~\cite{Alves:2005kr}.  This cross section
provides direct access to the isospin of $H^+$, allowing our doublet
model (in which $T^3_{H^+} = 1/2$) to be distinguished from the type-2
seesaw (in which the triplet state $\Phi^+$ has $T^3_{\Phi^+} = 0$).
We find that because of this isospin difference, the cross section for
$H^+H^-$ is 2.7 (2.6) times larger than that for $\Phi^+ \Phi^-$ for
$M_{H^+,\Phi^+} = 100$ (500)~GeV.~\footnote{For recent LHC
phenomenology studies of the triplet states in the type-2 seesaw, see
Refs.~\cite{Perez:2008ha,delAguila:2008cj}.  The signatures are
dominated by $\Phi^{++}\Phi^{--}$ and $\Phi^{\pm\pm} \Phi^{\mp}$
production.}

We note also that the branching fraction of $H^+$ to $\mu \nu$ or $e
\nu$ is always at least $1/3$ (Fig.~\ref{fig:brs}), which provides
distinctive search channels for $H^+$ at the LHC.  Details will be
given in a forthcoming paper~\cite{longpaper}.  Measurement of the
characteristic pattern of $H^+$ branching fractions would provide
strong evidence for the connection of $H^+$ to the neutrino sector, as
well as allowing a determination of the neutrino mass hierarchy and
providing some sensitivity to the lightest neutrino mass.

\subsection{Effects on the SM-like Higgs}

Finally we comment on the phenomenology of the SM-like Higgs $h^0$.
The charged Higgs will contribute to the one-loop amplitude for $h^0
\to \gamma\gamma$, yielding 
\begin{equation}
  \Gamma(h^0 \to \gamma\gamma) =
  \Gamma^{SM}(h^0 \to \gamma\gamma) \left[ 1 - \lambda_3 \delta \left[
  \frac{100~{\rm GeV}}{M_{H^+}} \right]^2 \right]^2,
\end{equation}
where for $M_h = 120$~GeV and $M_{H^+} = 100$ (200, 1000)~GeV, $\delta
= 0.20$ (0.17, 0.16).  This provides experimental access to the
coupling $\lambda_3$, which appears in the $h^0 H^+ H^-$, $h^0 A^0
A^0$, and $h^0 H^0 H^0$ vertices.  Again for $M_h = 120$~GeV, the
experimental precision with which $\Gamma(h^0 \to \gamma\gamma)$ can
be extracted at the LHC has been estimated at
$\sim$15--30\%~\cite{LHCcoups}; at the International Linear $e^+e^-$
Collider with 500~GeV center-of-mass energy the precision remains at
the $\sim$25\% level due to limited statistics~\cite{Desch:2003xq}.
This would provide sensitivity to $|\lambda_3| \sim 1$ only at the $1
\sigma$ level.  This precision could be improved to $\sim$5\% at a
Linear Collider with 1~TeV center-of-mass energy~\cite{Barklow:2003hz}
or $\sim$2\% at a photon collider running on the Higgs
resonance~\cite{photoncollider}, providing 4--10$\sigma$ sensitivity
to $|\lambda_3| \sim 1$, respectively.

More importantly, our model impacts the range of SM-like Higgs masses
allowed by the standard electroweak fit.  In particular, the new
scalars can increase the allowed range for $M_h$ by giving a positive
contribution to the $\rho$ parameter.  Using the result for a generic
two Higgs doublet model from Ref.~\cite{Haber:1992cn}, the new scalars
yield
\begin{equation}
  \Delta \rho = \alpha_{em} \Delta T 
  = \frac{\alpha_{em}}{8 \pi M_W^2 s_W^2} F(M_{H^+}^2, M_A^2),
\end{equation}
where $M_W$ is the $W$ boson mass, $s_W$ denotes the sine of the weak
mixing angle, and $F(m_1^2, m_2^2) \equiv (m_1^2 + m_2^2)/2 - [m_1^2
m_2^2/(m_1^2 - m_2^2)] \ln(m_1^2/m_2^2)$.  Here the shift in the
oblique parameter $\Delta T$ is defined relative to a SM reference
point with SM Higgs mass set equal to $M_h$.  It was shown in
Ref.~\cite{Barbieri:2006dq} that a SM-like Higgs mass in the range of
400--600~GeV can be made consistent with the electroweak fit if
$\Delta T \approx 0.25 \pm 0.1$.  This can be achieved in our model
for $M_{H^+} = 100$~GeV with $M_A \approx 200$--250~GeV, corresponding
to $\lambda_4 = 2 \sqrt{2} G_F (M_A^2 - M_{H^+}^2) \approx 1.0$--1.7.

For $M_h > 2 M_{H^+}$ or $2 M_A$, the additional decays $h^0 \to
H^+H^- \to \ell^+ \ell^{\prime -} \nu \bar \nu$ or $h^0 \to H^0 H^0,
A^0 A^0 \to 4 \nu$ appear.  The partial widths for these decays are
given above threshold by
\begin{eqnarray}
   \Gamma(h^0 \to H^+ H^-) &=& \frac{\lambda_3^2}{16 \pi \sqrt{2} G_F M_h}
   \sqrt{1 - \frac{4 M_{H^+}^2}{M_h^2}}, \nonumber \\
   \Gamma(h^0 \to A^0 A^0) &=& \Gamma(h^0 \to H^0 H^0) \nonumber \\
   &=& \frac{(\lambda_3 + \lambda_4)^2}{32 \pi \sqrt{2} G_F M_h}
   \sqrt{1 - \frac{4 M_A^2}{M_h^2}}.
\end{eqnarray}
For example, for $M_{H^+} = 100$~GeV and $M_h = 300$~GeV we obtain
$\Gamma(h^0 \to H^+H^-) = 3.0 \lambda_3^2$~GeV; for comparison the 
total width of a 300~GeV SM Higgs boson is 8.5~GeV~\cite{HDECAY}.

Below threshold, these Higgs-to-Higgs decays are suppressed by the
small $y_i^{\nu}$ unless $A^0,H^0 \to H^{\pm} W^{\mp}$ or $H^+ \to W^+
A^0/H^0$ is open.

\section{Conclusions}
\label{sec:conclusions}

We introduced a simple new TeV-scale model for Dirac neutrinos which
explains the smallness of neutrino masses by sourcing them from a
second Higgs doublet with tiny vev $\sim$~eV.
The model predicts distinctive decay patterns of $H^+$ controlled by
the neutrino mass spectrum and mixing matrix, which can be tested at
the LHC.  The isospin of $H^+$ can be measured at the LHC, allowing
the model to be distinguished from the type-2 seesaw for Majorana
neutrino mass generation.  The model also predicts a signal in $\mu
\to e\gamma$ at the currently-running MEG experiment if
$\sin\theta_{13} \gtrsim 0.01$ and $v_2 \lesssim 6$~eV.
Because the model conserves lepton number, neutrinoless double beta
decay is absent.

\begin{acknowledgments}
We thank Sacha Davidson, K.~Tsumura and P.~Watson for valuable
conversations and P.~Fileviez Perez for suggesting $h^0 \to
\gamma\gamma$.
This work was supported by the Natural Sciences and Engineering
Research Council of Canada.
\end{acknowledgments}


\end{document}